\documentclass{SciPost}

\usepackage{import}

\usepackage{orcidlink}
\usepackage{csquotes}
\usepackage{amsmath}
\usepackage{amsthm}
\usepackage{amssymb}
\usepackage{bm}
\usepackage{color}
\usepackage{ifthen}
\usepackage[svgnames]{xcolor}
\usepackage{multirow,bigdelim}
\usepackage{comment}
\usepackage{url}
\usepackage{soul}
\usepackage{hyperref}

\hypersetup{
    colorlinks,
    linkcolor={red!50!black},
    citecolor={blue!50!black},
    urlcolor={blue!80!black}
}

\usepackage[capitalize]{cleveref}

\usepackage[bitstream-charter]{mathdesign}
\DeclareSymbolFont{usualmathcal}{OMS}{cmsy}{m}{n}
\DeclareSymbolFontAlphabet{\mathcal}{usualmathcal}

\fancypagestyle{SPstyle}{
\fancyhf{}
\lhead{\colorbox{scipostblue}{\bf \color{white} ~SciPost Physics }}

\fancyfoot[C]{\textbf{\thepage}}
}

\newcommand{\1}{\mbox{1}\hspace{-0.25em}\mbox{l}}

\newcommand{\ii}{\mathrm{i}}

\newcommand{\tableblue}[1]{{\textcolor{blue}{#1}}}

\begin{document}

\pagestyle{SPstyle}

\begin{center}{\Large \textbf{\color{scipostdeepblue}{
Hopf Exceptional Points
}}}\end{center}

\begin{center}\textbf{
Tsuneya Yoshida\orcidlink{0000-0002-2276-1009}\textsuperscript{1, 2$\star$}, Emil J. Bergholtz\orcidlink{0000-0002-9739-2930}\textsuperscript{3$\dagger$} and 
Tom\'a\v{s} Bzdu\v{s}ek\,\orcidlink{0000-0001-6904-5264}\textsuperscript{4$\ddagger$}
}\end{center}

\begin{center}
{\bf 1} Department of Physics, \href{https://ror.org/02kpeqv85}{Kyoto University}, Kyoto 606-8502, Japan
\\
{\bf 2} Institute for Theoretical Physics, \href{https://ror.org/05a28rw58}{ETH Zurich}, 8093 Zurich, Switzerland
\\
{\bf 3} Department of Physics, \href{https://ror.org/05f0yaq80}{Stockholm University}, \\ AlbaNova University Center, 10691 Stockholm, Sweden
\\
{\bf 4} Department of Physics, \href{https://ror.org/02crff812}{University of Z\"urich}, \\ Winterthurerstrasse 190, 8057 Z\"urich, Switzerland
\\
%
%
[\baselineskip]
$\star$ \href{mailto: yoshida.tsuneya.2z@kyoto-u.ac.jp}{\small yoshida.tsuneya.2z@kyoto-u.ac.jp}\,,\quad
$\dagger$ \href{mailto: emil.bergholtz@fysik.su.se}{\small emil.bergholtz@fysik.su.se}\,,\quad
$\ddagger$ \href{mailto: tomas.bzdusek@uzh.ch}{\small tomas.bzdusek@uzh.ch}
\end{center}

\section*{\color{scipostdeepblue}{Abstract}}
\textbf{%
Exceptional points at which eigenvalues and eigenvectors of non-Hermitian matrices coalesce are ubiquitous in the description of a wide range of platforms from photonic or mechanical metamaterials to open quantum systems.
Here, we introduce a class of \emph{Hopf exceptional points} (HEPs) that are protected by the Hopf invariants (including the higher-dimensional generalizations) and which exhibit phenomenology sharply distinct from conventional exceptional points. 
Saliently, owing to their $\mathbb{Z}_2$ topological invariant related to the Witten anomaly, three-fold HEPs and symmetry-protected five-fold HEPs act as their own \enquote{antiparticles}.
Furthermore, based on higher homotopy groups of spheres, we predict the existence of multifold HEPs and symmetry-protected HEPs with non-Hermitian topology captured by a range of finite groups (such as $\mathbb{Z}_3$, $\mathbb{Z}_{12}$, or $\mathbb{Z}_{24}$) beyond the periodic table of Bernard-LeClair symmetry classes. 
}

\vspace{\baselineskip}

\vspace{10pt}
\noindent\rule{\textwidth}{1pt}
\tableofcontents
\noindent\rule{\textwidth}{1pt}
\vspace{10pt}


\section{Introduction}

The notion of topology plays a pivotal role in modern condensed matter physics of both quantum~\cite{Hasan_TIRev_RMP2010,Qi_TIRev_RMP2010,Kane_Z2TI_PRL05_1,Kane_Z2TI_PRL05_2,Qi_TFT_PRB08} and classical systems~\cite{Haldane_TopoPhoto_PRL2008,Lu_TopoPhoto_NatPhoto2014,Prodan_TopoMech_PRL2009,Albert_TopoEle_PRL2015,Lee_Topoele_CommPhys2018}. 
Saliently, topological physics implies the possibility of various exotic quasiparticles~\footnote{
Here, the term ``quasiparticle" denotes a topologically protected band singularity, as is widely used in the literature~\cite {Hasan_TIRev_RMP2010,Qi_TIRev_RMP2010,Bradlyn_MFermi_Science2016}.
}.
One of the prime examples is a Majorana zero mode~\cite{Kitaev_chain_01,Moore_Majorana_NPhysB1991,Read_Majorana_PRB2000,Fu_Majorana_PRL2008} whose antiparticle is itself.
In addition, a Weyl fermion~\cite{Weyl_WeylFermi_ZPhys1929,Murakami_Weyl_NJP2007,Wan_WSM_RB2011,Burkov_WSM_PRL2011,Yan_WSMReview_AnnRevCondMattPhys2017,Xu_WSMTaAsExp_Science2015,Lv_WSMTaAsExp_PRX2015}, protected by Chern number in topological semimetals, is a source of negative magnetoresistance~\cite{Huang_TaAsNegMagReg_PRX2015} which is a signal of chiral anomaly.
The periodic table for the ten Altland-Zirnbauer symmetry classes provides a systematic understanding of the topological obstructions inducing these exotic excitations in Hermitian systems~\cite{Schnyder_classification_free_2008,Kitaev_classification_free_2009,Ryu_classification_free_2010,Shiozaki_class_PRB2014,Morimoto_TSMclass_PRB2014,Chiu_TSMclass_PRB2014,Bzdusek_TSMclass_PRB2017}.

Notably, open systems coupled to environments host topological excitations for which non-Hermiticity is essential~\cite{Bergholtz_nH_RMP19,Ashida_nHReview_AdvPhys2020}, such as exceptional points~\cite{TKato_EP_book1966,Heiss_EPopenq_PRE1998,Rotter_EP_JPA09,Berry_EP_CzeJPhys2004,Heiss_EP_JPA12}. 
At exceptional points, two energy bands touch in both the real and the imaginary parts. 
Such band touchings are protected by the winding topology of energy eigenvalues~\cite{Shen_TopBandEP_PRL2018}. 
In sharp contrast to Hermitian topological excitations, exceptional points exhibit a dispersion with a fractional exponent. 
Exceptional points and their variants~\cite{Xu_EP2L_PRL2017,Carlstrom_EP2Knot_PRA2018,Budich_SPERs_PRB19,Okugawa_SPERs_PRB19,Zhou_SPERs_Optica19,Yoshida_SPERs_PRB19,Kimura_SPES_PRB2019,Yang_EPdoubling_PRL2021,Denner:2021,Yoshida_EPDiscInd_PRB2022,Shen_TopBandEP_PRL2018,Sun:2020,Wojcik_EP_PRB2020,Li_nHHomotopy_PRB2021,Yang_nHHomotopy_RepProgPhys2024} are reported for a wide range of platforms from quantum systems~\cite{VKozii_nH_arXiv17,Zyuzin_nHEP_PRB18,Yoshida_EP_DMFT_PRB18,Yoshida_nHReview_PTEP20,Shen_QOsci_PRL2018,Nagai_EPDMFT_PRL2020,Shallem_LEP_NJP2015,Miganti_LEP_PRA2019,Nakagawa_LHubb_PRL2021,Chen_EPqubit_PRL2022} to metamaterials~\cite{Dembowski_EPPhoto_PRL2001,Lee_EPPhoto_PRL2009,Doppler_EPPhono_Nature2016,Zhen_EPPhoto_Nature2015,KTakata_EP_PRL2018,Zhou_EPPhoto_Science2018,Meng_EPPhoto_ApplPhysLett2024,Ozawa_TopoPhoto_RMP2019,AliMiri_EPOptoPhoto_Science2019,Zhu_EPPhonon_PRX2014,Shi_EPPhono_NatComm2016,Yoshida_SPERs_mech19,HopfMan_ResSkinEleCir_PRRes2020,Li_EPDiff_Science2019}, indicating the ubiquity of these non-Hermitian excitations.
In particular, the high controllability of synthetic systems allows for the realization of exceptional points in dimensions larger than three~\cite{Dembowski_EPPhoto_PRL2001,Lee_EPPhoto_PRL2009,Tang_EPn_Science2020,Tang_EP3Mech_NatComm2023}.

Among the various exceptional points, multifold exceptional points exhibit an $n$-fold band touching~\cite{Heiss_EPn_JPA2008,Graefe_EPn_JPA2008,Heiss_EP_JPA12,Demange_EP3MathPhoto_JPhysA2012,JWRyu_EPn_PRA2022,Yang_EPn_PRB2023}
whose stability due to topology and symmetry has recently been revealed in Refs.~\cite{Delplace_EP3_PRL2021,Mandal_EP3_PRL21,Sayyad_EPn_PRR2022,Montag_EPn_PRR2024,Yoshida_EPn_PRR2024,Stalhammar_EPn_arXiv2024}.
These unique excitations are beyond the existing classification table for Bernard-LeClair symmetry classes~\cite{Kawabata_EPClass_PRL19,Gong_class_PRX18,KKawabata_TopoUni_NatComm19,Kawabata_gapped_PRX19,Zhou_gapped_class_PRB19} and discussed in interdisciplinary fields~\cite{Lin_EP3phtonic_RPL2016,Schnabel_EP3PTPhotonic_PRA2017,Wiersig_EPn_PhotoRes2020,Wiersig_EP5photo_PRA2022,Wang_EPnPTPhotoSciAdv2023,Hodaei_EPnPT_Nature2017,Zhou_EPPhoto_Science2018,Tang_EPn_Science2020,Bai_NLEP3_PRL2023,Liu_EPn_PRR2023,Patli_EPnPT_Nature2022,Leclerc_EP4star_PRR2024,Crippa_EP4Corr_PRB2021,MarcusClassification,Hatano_EP3Lindblad_MolPhys2019,Khandelwal_EP3Lindblad_PRXQ2021,Wu_EP3openQ_NatNano2024,Ryu_EP4_PRA2022}.
However, topology of the formerly reported $n$-fold exceptional points (EP$n$s) is generally characterized by $\mathbb{Z}$ invariants ~(i.e., winding topology) for $n\geq 3$~\cite{Delplace_EP3_PRL2021,Yoshida_EPn_PRR2024}.
This, in particular, implies that an exceptional point cannot be its own \enquote{antiparticle}.
If such $n$-fold band touching exists, it should be distinguished from the formerly reported EP$n$.

\begin{figure}[!t]
\begin{center}
\includegraphics[width=0.6\hsize,clip]{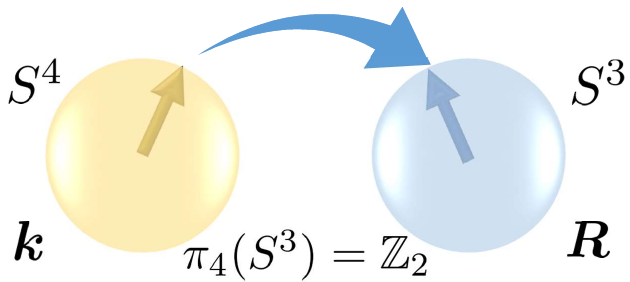}
\end{center}
\caption{
Illustration of $\mathbb{Z}_2$ topology protecting a three-fold Hopf exceptional point (HEP3).
The resultant vector $\bm{R}(\bm{k})$ defines a map from a 4-sphere in the momentum (or parameter) space to a 3-sphere in the space of the resultant vector [see~\cref{eq: def r_j,eq: Rvec Generic 3x3}]. 
If the map is topologically nontrivial, an HEP3 with $\mathbb{Z}_2$ topology exists inside the 4-sphere. 
}
\label{fig:Sketch_S4toS3}
\end{figure}

In this work, we report novel non-Hermitian topological excitations, dubbed $n$-fold Hopf exceptional points 
(HEP$n$s, $n\,{=}\,3,\ldots,7$)
\footnote{
To achieve concise terminology, we refer to \emph{all} such EPs protected by topologically nontrivial maps between higher-dimensional spheres with $d> m \geq 2$ collectively as \emph{Hopf exceptional points}, although not all of these maps are termed as Hopf maps in the mathematical literature.
}
,
which are topologically stabilized by higher homotopy groups of spheres, labeled $\pi_d(S^m)$, with integers $d \,{>}\, m \,{\geq}\, 2$.
Saliently, an HEP$3$ and a symmetry-protected HEP$5$ exhibit unusual $\mathbb{Z}_2$ topology, meaning that they act as their own antiparticle. 
We trace this striking feature to the homotopy group $\pi_4(S^3)\,{=}\,\mathbb{Z}_2$ (see \cref{fig:Sketch_S4toS3}) classifying the map of resultants. 
We further discover symmetry-protected HEP$4$ whose topology is classified by $\pi_3(S^2)\,{=}\,\mathbb{Z}$.
By systematically leveraging higher homotopy groups of spheres, we elucidate the potential presence of HEP$n$s characterized by abundant finite groups (e.g., $\mathbb{Z}_3$, $\mathbb{Z}_{12}$, and $\mathbb{Z}_{24}$) beyond the existing classification table.

\section{HEP3s with \texorpdfstring{$\mathbb{Z}_2$}{Z2} topology}
As a first example, we consider a generic three-band non-Hermitian Hamiltonian $H(\bm{k})$ in a five-dimensional momentum (or parameter) space denoted by $\bm{k}\,{=}\, (k_1,\ldots,k_5)$. 
Such a momentum space whose dimensionality is higher than three can be accessed in metamaterials~\cite{Dembowski_EPPhoto_PRL2001,Lee_EPPhoto_PRL2009,Tang_EPn_Science2020,Tang_EP3Mech_NatComm2023,Yoshida_EPn_PRR2024}. 
The formation of an EP3 is captured by vanishing resultants $r_j\,{\in}\,\mathbb{C}$
\begin{align}
\label{eq: def r_j}
r_j(\bm{k})&=\mathrm{Res}[\partial^{n-1-j}_E P(E,\bm{k}),\partial^{n-1}_E P(E,\bm{k})],
\end{align}
with $j\,{=}\,1,2,\ldots,n\,{-}\,1$ and $n\,{=}\,3$ (see \cref{appsec: Res})
~\footnote{
The vanishing resultants $r_j=0$ ($j=1,2,\ldots,n-1$) indicate the $n$-tiple root; it captures the algebraic multiplicity rather than geometric multiplicity.
We note, however, that these two coincide with each other because breaking the coincidence requires fine-tuning.
}
~\footnote{
For $n=2$, \cref{eq: def r_j} reduces to the discriminant whose winding topology protects EP2 for systems without symmetry~\cite{Shen_TopBandEP_PRL2018,Yang_EPdoubling_PRL2021}.
}
.
Here, $P(E,\bm{k})\,{=}\,\mathrm{det}[H(\bm{k})\,{-}\,E\1]\,{\in}\,\mathbb{C}$ is the characteristic polynomial,  and $\partial_E$ denotes derivative with respect to the polynomial variable $E\,{\in}\,\mathbb{C}$.

To expose whether the EP$3$ is in fact an HEP$3$, we consider a 4-sphere $S^4 \subset \mathbb{R}^5$ enclosing the band touching. Assuming $(r_1,r_2)\neq (0,0)$ on the $4$-sphere, we introduce a normalized vector $\bm{n}\,{=}\,\bm{R}/\lVert\bm{R}\rVert$ ($\lVert\bm{R}\rVert\,{=}\,\sqrt{\bm{R}\cdot\bm{R}}$) with the resultant vector
\begin{align}
\label{eq: Rvec Generic 3x3}
 \bm{R}(\bm{k}) &= (\mathrm{Re}[r_1],\mathrm{Im}[r_1],\mathrm{Re}[r_2],\mathrm{Im}[r_2]).
\end{align}
Vector $\bm{n}(\bm{k})$ defines a map from $S^4$ to $S^3$ whose topology is classified by an element of the homotopy group $\pi_4(S^3)\,{=}\,\mathbb{Z}_2$.  
When the map of $\bm{n}(\bm{k})$ possesses nontrivial $\mathbb{Z}_2$ topology, the enclosed band touching is an HEP3. 
In passing, we note that the resultant topology is neither point-gap nor line-gap topology~\cite{Delplace_EP3_PRL2021,Yoshida_EPn_PRR2024} (see also \cref{appsec: ResW} for a brief overview of the resultant winding topology).

The $\mathbb{Z}_2$ invariant $\nu_{\mathrm{F}}$ of maps from $S^4$ to $S^3$, originally discovered by Freudenthal~\cite{Freudenthal:1938}, was considered in physics in the context of the Witten anomaly in $\mathrm{SU(2)}$ gauge theory~\cite{Witten_Z2inv_NuclPhys1983} and to describe topological defects in superfluid ${}^3$He~\cite{Grinevich_JLTP1988}.
Numerical computation of this $\mathbb{Z}_2$ invariant is carried out by the following representation~\cite{Schuster_FHI_PRL2019}
\begin{align}
\label{eq: Z2 inv}
\nu_{\mathrm{F}} &= \frac{1}{4\pi}
\oint d^4p \epsilon^{\mu\nu\rho\lambda} [\partial_{\mu}
\Delta\varphi(\bm{p})]
A_{\nu}F_{\rho\lambda}
\end{align}
with $\mu,\nu,\rho,\lambda\,{=}\,1,\ldots,4$ and anti-symmetric tensor $\epsilon^{\mu\nu\rho\lambda}$ taking $\epsilon^{1234}\,{=}\,1$.
The Berry connection $A_{\mu}$ and the Berry curvature $F_{\mu\nu}$ are obtained from the \emph{resultant Hamiltonian}, and the phase $\Delta\varphi(\bm{p})$ ($0 \leq \Delta \varphi \leq 2\pi$) is obtained from the resultant vector. 
For the precise definitions of $A_{\mu}$, $F_{\mu\nu}$ and $\Delta\varphi$, see \cref{appsec: phi A F}. Vector $\bm{p}$ parametrizes the $4$-sphere in the momentum space. 
While the integral in \cref{eq: Z2 inv} can take an arbitrary integer value, gauge transformations can change $\nu_{\mathrm{F}}$ by multiples of two~\cite{Witten_Z2inv_NuclPhys1983}.

We demonstrate the emergence of an HEP3 in five dimensions by analyzing a toy model.  
The Hamiltonian reads
\begin{align}
\label{eq: toy Z2EP3}
H &=
\left(
\begin{array}{ccc}
0 & 1 & 0  \\
0 & 0 & 1  \\
\frac{\zeta_2}{6} & \zeta_1 & 0  \\
\end{array}
\right),
\end{align}
where the functions $\zeta_{1,2}(\bm{k})$ are parameterized by $\delta$, $m_0$, and $f(k_5)$ (for the explicit form, see \cref{appsec: toy Z2EP3}). 
Here the function $f(k_5)$ is either even [$f(k_5)\,{=}\,1$] or odd [$f(k_5)\,{=}\,2\sin (k_5/2)$].

\begin{figure}[!t]
\begin{center}
\includegraphics[width=0.75\hsize,clip]{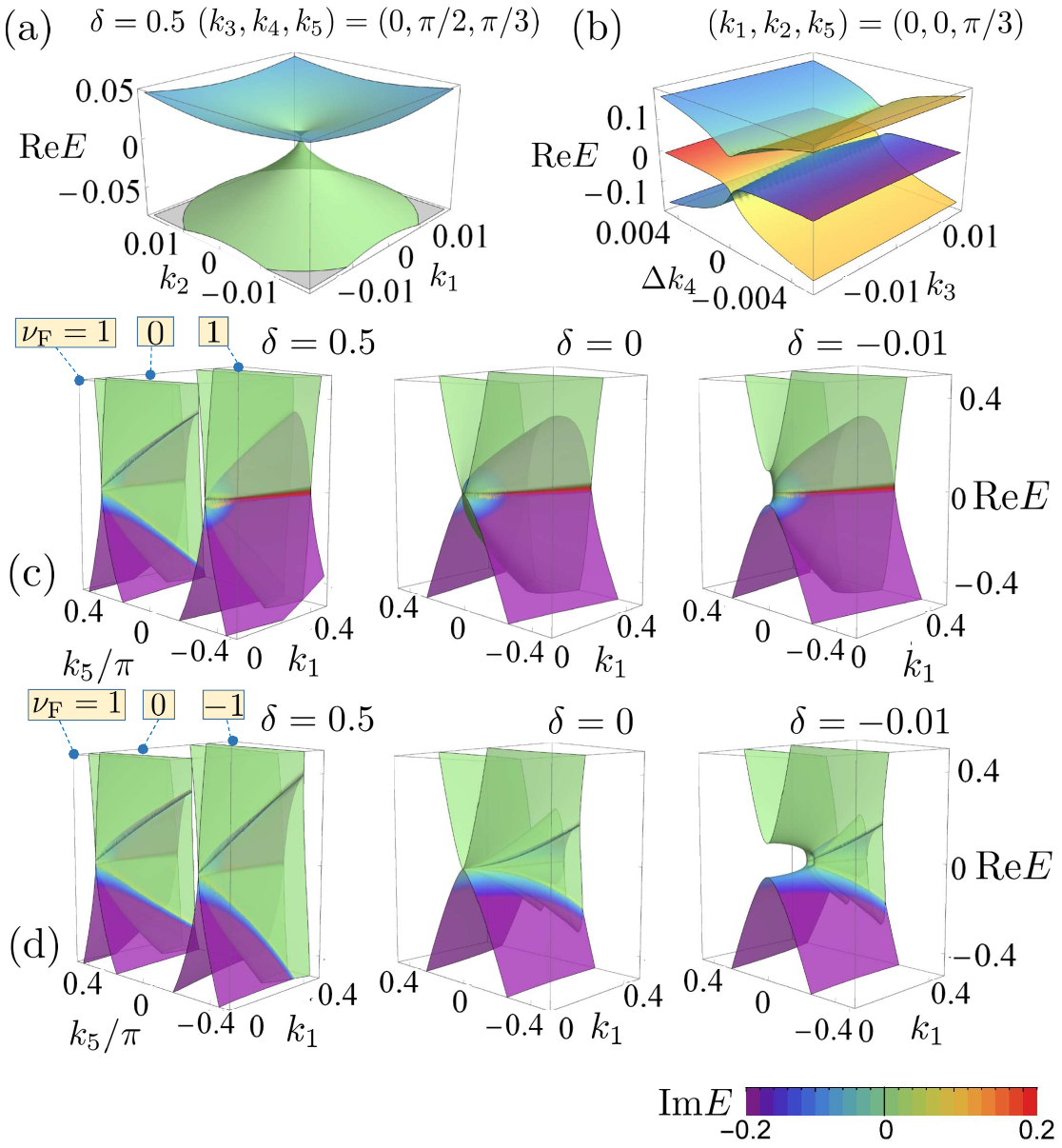}
\end{center}
\caption{
Energy bands of Hamiltonian~\eqref{eq: toy Z2EP3} for $m_0\,{=}\,1.5$.
The real (imaginary) part is represented as height (color).
In panel~(a), the complex conjugate of the upper band is omitted.
Panels~(a) and (b) are obtained for $k_5\,{=}\,\pi/3$, $\delta\,{=}\,0.5$, and $f(k_5)\,{=}\,1$. 
Here, $\Delta k_4$ is defined as $\Delta k_4\,{=}\,k_4-\pi/2$.
Panel~(c) [(d)] displays pair annihilation of HEP3s for $(k_2,k_3,k_4)\,{=}\,(0,0,\pi/2)$, and $f(k_5)\,{=}\,1$ [$f(k_5)\,{=}\,2\sin (k_5/2)$].
In these panels, numerically computed $\nu_\mathrm{F}$ for $k_5\,{=}\,-\pi/2$, $0$, and $\pi/2$ at $\delta\,{=}\,0.5$ is represented by numbers highlighted in yellow.
We used a mesh of $40^4$ points to evaluate the integrals with momenta $k_{1,2,3}\in[-\pi,\pi]$ and $k_4 \in [0,2\pi]$. 
}
\label{fig: Z2Ep3 Bands}
\end{figure}

To motivate the Hamiltonian form in Eq.~\eqref{eq: toy Z2EP3}, recall that a three-fold exceptional point is generally captured by a $3 \times 3$ Jordan block form~\cite{Sayyad_EPn_PRR2022}.
The Hamiltonian including the perturbations $\zeta$'s in the bottom row
ensures that the resultants [recall Eq.~\eqref{eq: def r_j}] are expressed as $r_j \propto \zeta_j$ for $j\in \{1,2\}$~\cite{Yoshida_EPn_PRR2024}, which provides clear correspondence between the perturbations $\zeta$'s and the Hopf topology of $\bm{R}$.
The sought higher Hopf topology is then ensured by imposing the appropriate dependence~\cite{Schuster_FHI_PRL2019} of the functions $\zeta_{1,2}(\boldsymbol{k})$ on momenta (parameters) $\boldsymbol{k}$.~\footnote{
Another choice of $\zeta$'s yields an EP3 with winding topology~\cite{Yoshida_EPn_PRR2024} (see Appendix~\ref{appsec: ResW}).
}
The vanishing resultant vector determines where the HEP3s emerge in the momentum space.
The explicit form of the resultant vector [see \cref{appsec: toy Z2EP3}] indicates the emergence of HEP3s at $\bm{k}\,{=}\,(0,0,0,\pi/2, \pm \pi/3)$ for $\delta\,{=}\,0.5$ and $m_0\,{=}\,1.5$. Figure~\ref{fig: Z2Ep3 Bands}(a,b) displays the emergence of the HEP3 at $\bm{k}\,{=}\,(0,0,0,\pi/2, \pi/3)$.

Notably, the HEP acts as its own antiparticle as a direct consequence of its $\mathbb{Z}_2$ topology~\footnote{
The pair annihilation of topological nodes with the same charge is commonly observed for $\mathbb{Z}_2$ topology; for instance, Ref.~\cite{Yoshida_EPDiscInd_PRB2022} demonstrates pair annihilation of EP2s with $\mathbb{Z}_2$ topology.
}.
This fact is elucidated by examining pair annihilation of HEP3s in two cases: $f(k_5)\,{=}\,1$ and $f(k_5)\,{=}\,2\sin(k_5/2)$ [see \cref{fig: Z2Ep3 Bands}(c,d)].
For $f(k_5)\,{=}\,1$ and $\delta\,{=}\,0.5$, the system hosts two HEP3s demarcated by the planes at $k_5\,{=}\,\pi/2$, $0$ and $-\pi/2$ where numerically computed $\nu_{\mathrm{F}}$ is equal to $1$, $0$, and $1$, respectively [for computation of $\nu_{\mathrm{F}}$, see \cref{appsec: toy Z2EP3}]. 
As $\delta$ decreases, the two HEP3s approach and annihilate each other [see \cref{fig: Z2Ep3 Bands}(c)].
Changing the parity of $f(k_5)$ flips the sign of the numerically computed $\nu_{\mathrm{F}}$ at $k_5\,{=}\,-\pi/2$. 
Even in this case, pair annihilation occurs [see \cref{fig: Z2Ep3 Bands}(d)].
The occurrence of pair annihilation in both arrangements manifests that HEP3s are indeed protected by $\mathbb{Z}_2$ topology, implying that an HEP3 is its own antiparticle.

\section{Symmetry-protected HEP5s with \texorpdfstring{$\mathbb{Z}_2$}{Z2} topology}

Symmetry further enriches Hopf exceptional points, as exemplified by the emergence of symmetry-protected HEP5 in five dimensions.
We consider a five-band non-Hermitian Hamiltonian which preserves parity-time ($PT$-) symmetry
\begin{align}
\label{eq: pT symm}
 U_{\mathrm{PT}}^{\phantom{-1}} H^*(\bm{k}) U^{-1}_{\mathrm{PT}} &= H(\bm{k})
\end{align}
with a unitary matrix satisfying $U_{\mathrm{PT}}^{\phantom{*}}U^*_{\mathrm{PT}}\,{=}\,\1$. 
Here, $\1$ is the identity matrix, and asterisks denote complex conjugation. 
The $PT$-symmetry imposes the constraint 
\begin{align}
\label{eq: PT-symm PE}
 P(E) &= P^*(E^*),
\end{align}
indicating that all coefficients of the characteristic polynomial $P(E)$ are real at each $\boldsymbol{k}$.
For $n=5$, the formation of an EP5 is captured by vanishing resultants $r_{1,\ldots,4}\,{=}\,0$ [see \cref{eq: def r_j}].
Due to $PT$-symmetry [see \cref{eq: pT symm,eq: PT-symm PE}] these resultants are real: $r_{1,\ldots,4}\in \mathbb{R}$.

To expose whether the EP5 is in fact an HEP5, we consider a 4-sphere $S^4 \subset \mathbb{R}$ enclosing the band touching.
Assuming $r_{1,\ldots,4} \neq 0$ on the 4-sphere, we introduce a normalized 
vector $\bm{n}\,{=}\,\bm{R}/\lVert\bm{R}\rVert$ with 
\begin{align}
\label{eq: Rvec PT 5x5}
 \bm{R} &= (r_1,r_2,r_3,r_4)^{\mathrm{T}}.
\end{align}
The normalized vector $\bm{n}$ defines a map from $S^4$ to $S^3$ whose topology is classified by $\pi_4(S^3)\,{=}\,\mathbb{Z}_2$.
When the map of $\bm{n}$ possesses nontrivial $\mathbb{Z}_2$ topology, the enclosed band touching is a symmetry-protected HEP5.
The topological invariant is introduced in a similar way as \cref{eq: Z2 inv} with the only difference that we compute the $\mathbb{Z}_2$ invariant from the resultant vector in \cref{eq: Rvec PT 5x5} instead of the one in \cref{eq: Rvec Generic 3x3}.

\begin{figure}[!t]
\begin{center}
\includegraphics[width=0.75\hsize,clip]{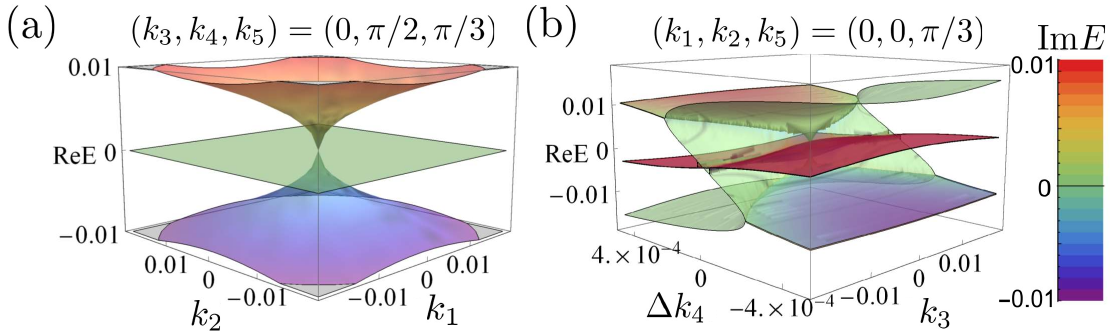}
\end{center}
\caption{
Energy bands of Hamiltonian~\eqref{eq: toy Z2 EP5} for $f(k_5)\,{=}\,1$, $k_5\,{=}\,\pi/3$, $m_0\,{=}\,1.5$ and $\delta\,{=}\,0.5$. 
The real (imaginary) part is represented as height (color).
Panel~(a) [(b)] displays the data for $(k_3,k_4)\,{=}\,(0,\pi/2)$ [$(k_1,k_2)\,{=}\,(0,0)$].
In panel~(a), the complex conjugates of the upper and the lower bands are omitted. In panel~(b), $\Delta k_4$ is defined as $\Delta k_4\,{=}\,k_4-\pi/2$.
}
\label{fig: Z2 PTEp5 Bands}
\end{figure}

We demonstrate the emergence of a symmetry-protected HEP5 by analyzing a toy model whose Hamiltonian reads
\begin{align}
\label{eq: toy Z2 EP5}
H &= 
\left(
\begin{array}{ccccc}
0 & 1 & 0 & 0 & 0 \\
0 & 0 & 1 & 0 & 0 \\
0 & 0 & 0 & 1 & 0 \\
0 & 0 & 0 & 0 & 1 \\
\frac{\zeta_4}{(5!)^3} & \frac{\zeta_3}{(5!)^2} & \frac{\zeta_2}{5!2!} & \frac{\zeta_1}{3!} &0 \\
\end{array}
\right)
\end{align}
with real functions $\zeta_{1,\ldots,4}(\bm{k})$ parameterized by $\delta$, $m_0$, and $f(k_5)$ [for the explicit form, see \cref{appsec: toy Z2EP5}]. 
Here, $f(k_5)$ is either $f(k_5)\,{=}\,1$ or $f(k_5)\,{=}\,2\sin (k_5/2)$. 
This Hamiltonian preserves $PT$-symmetry [Eq.~\eqref{eq: pT symm}] with $U_{\mathrm{PT}}\,{=}\,\1$.
In analogy with the discussion of the Hamiltonian in \cref{eq: toy Z2EP3}, the perturbation of the $5\times 5$ Jordan form in \cref{eq: toy Z2 EP5} is chosen so that the resultants obey $r_j \propto \zeta_j$ for each $j\in\{1,\ldots,4\}$.
The explicit form of $\zeta_j(\boldsymbol{k})$ follows from a known representative of the nontrivial class in $\pi_4(S^3)$~\cite{Schuster_FHI_PRL2019}.

The vanishing resultant vector $\bm{R}\propto ( \zeta_1,\zeta_2,\zeta_3,\zeta_4 )^\mathrm{T}\,{=}\,\bm{0}$ specifies where the symmetry-protected HEP5s emerge.
The explicit form of the resultant vector [see \cref{appsec: toy Z2EP5}] indicates that the model in \cref{eq: toy Z2 EP5} hosts symmetry-protected $\mathbb{Z}_2$ HEP5s at $\bm{k}\,{=}\,(0,0,0,\pi/2,\pm \pi/3)$ which possesses $\mathbb{Z}_2$ topological charge [see \cref{fig: Z2 PTEp5 Bands}].

The argument of $PT$-symmetry protected HEPs can be applied to other cases of symmetry: pseudo-Hermiticity, $CP$-, and chiral symmetry.
We consider a Hamiltonian preserving pseudo-Hermiticity
\begin{align}
\label{eq: pseudoH}
U_{\mathrm{pH}} H^\dagger(\bm{k}) U^\dagger_{\mathrm{pH}}
 &= H(\bm{k})
\end{align}
with $U^\dagger_{\mathrm{pH}}$ being a unitary matrix and dagger denotes Hermitian conjugation.
Because the transposition does not affect the determinant, \cref{eq: pseudoH} leads to \cref{eq: PT-symm PE}, implying that the resultants [\cref{eq: def r_j}] are real.

For $CP$- and chiral symmetry, the Hamiltonian obeys 
\begin{align}
\label{eq: CP and chiral}
U_{\mathrm{CP}} H^*(\bm{k}) U^\dagger_{\mathrm{CP}}
 &=-H(\bm{k}), \\
 U_{\mathrm{\Gamma}} H^\dagger(\bm{k}) U^\dagger_{\mathrm{\Gamma}} &=
 -H(\bm{k})
\end{align}
with $U_{\mathrm{CP}}$ and $U_{\mathrm{\Gamma}}$ being unitary matrices. 
In these cases, replacing $H$ to $H'\,{=}\,iH$ reduces to the case of $PT$-symmetry or pseudo-Hermiticity. 
Therefore, symmetry-protected HEP$n$s may emerge when systems preserve pseudo-Hermiticity, $CP$-, or chiral symmetry.

\section{Symmetry-protected HEP4s with \texorpdfstring{$\mathbb{Z}$}{Z} topology}

Symmetry protection also enables HEP$4$s with $\mathbb{Z}$ topology.
To illustrate such a possibility, we consider a four-band non-Hermitian Hamiltonian with $PT$-symmetry [see \cref{eq: pT symm}] in a four-dimensional momentum space described by $\bm{k}\,{=}\,(k_1,k_2,k_3,k_4)$.
For $n=4$, the formation of an EP4 is captured by vanishing resultants $r_{1,\ldots,3}\,{=}\,0$ [see \cref{eq: def r_j}].

\begin{figure}[!t]
\begin{center}
\includegraphics[width=0.75\hsize,clip]{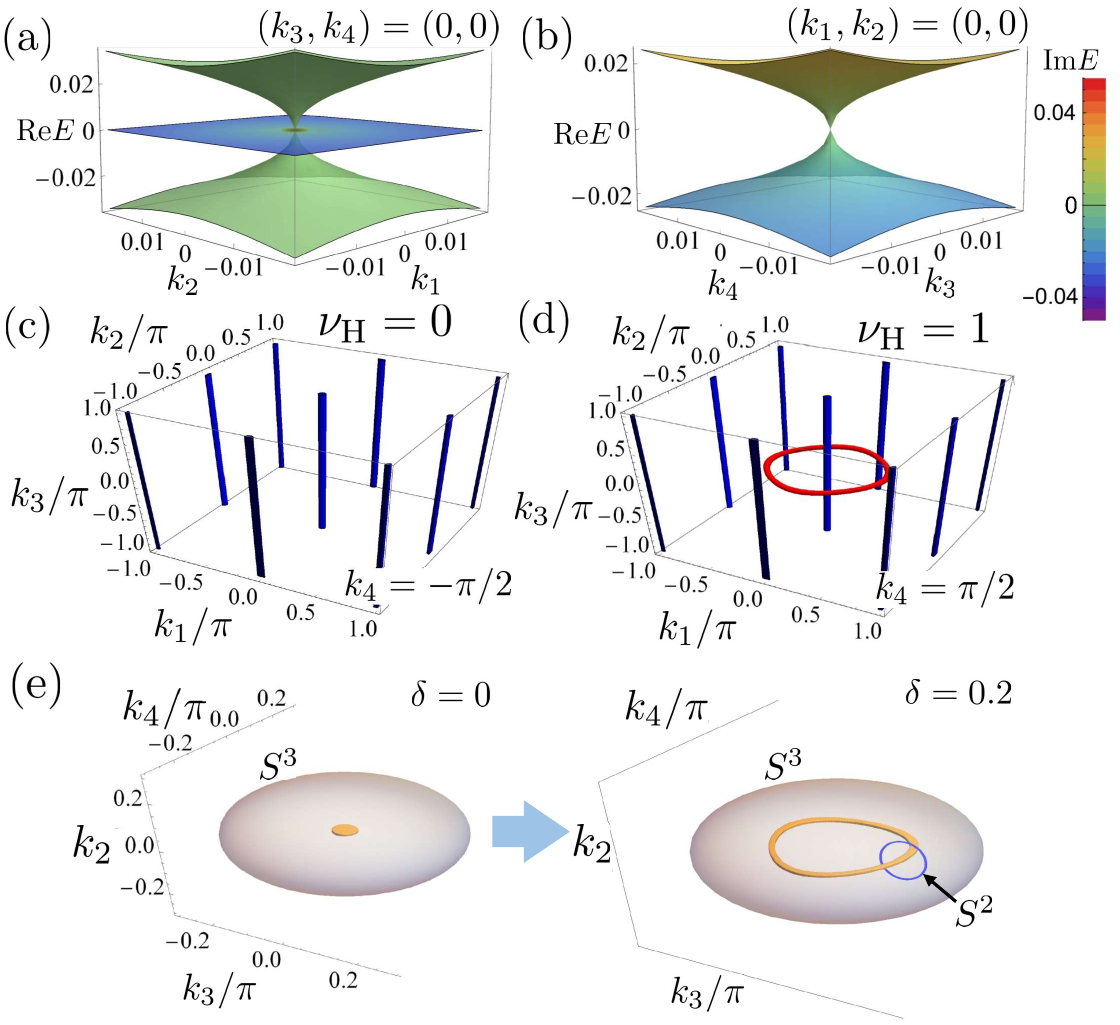}
\end{center}
\caption{
(a) and (b):~Energy eigenvalues of Hamiltonian in \cref{eq: toy Hopf EP4} for $m_0\,{=}\,3$ and $\delta\,{=}\,0$. 
The real (imaginary) part is represented as height (color). 
In panel~(b), complex conjugates of the upper and lower bands are omitted. 
(c) [(d)]:~Lines in the momentum space $(k_1,k_2,k_3)$ for $m_0\,{=}\,3$, $\delta\,{=}\,0$ and $k_4\,{=}\,-\pi/2$ [$k_4\,{=}\,\pi/2$], where red and blue lines denote the momenta satisfying $\bm{R}\,{\propto}\,(0,0,1)^{\mathrm{T}}$ and $\bm{R}\,{\propto}\,(0,0,-1)^{\mathrm{T}}$, respectively.
The linking of these lines determines the value of the Hopf invariant $\nu_{\mathrm{H}}$.
(e):~Momenta satisfying $\bm{R}(\bm{k})\,{=}\,\bm{0}$ for $k_1\,{=}\,0$ and $\delta\,{=}\,0$~resp.~$\delta\,{=}\,0.2$ (orange manifolds). 
As $\delta$ is introduced, the symmetry-protected HEP4 inflates into a loop. 
The gray oval and the blue loop illustrate the $S^3$ resp.~$S^2$, both extending in the fourth dimension $k_1$ (not shown), on which one computes the Hopf invariant $\nu_{\mathrm{H}}$ resp.~the resultant winding number $W_{2}$ (see \cref{appsec: ResW})~\cite{Yoshida_EPn_PRR2024}.
}
\label{fig: HopfEP4_Ek_inv}
\end{figure}

To expose whether the EP4 is in fact an HEP4, we consider a $3$-sphere $S^3\subset \mathbb{R}^4$ enclosing the band touching. 
The homotopy group $\pi_3(S^2)\,{=}\,\mathbb{Z}$ implies the existence of nontrivial maps $\bm{n}(\bm{k})\,{=}\,\bm{R}/\lVert \bm{R}\rVert$. 
When the map of $\bm{n}$ possesses nontrivial $\mathbb{Z}$ topology, the enclosed band touching is a symmetry-protected HEP4.

The topology of such HEP$4$s is characterized by the Hopf invariant, expressed as~\cite{Moore_HopfIns_PRL2008,Deng_HopfIns_PRB2013,Yang_HopfEP_PRB2019}
\begin{align}
\label{eq: Hopf inv}
\nu_{\mathrm{H}}= \oint \frac{d^3\bm{p}}{2}\epsilon^{\mu\nu\rho}A_{\mu}F_{\nu\rho},
\end{align}
with $\mu,\nu,\rho=1,2,3$.
Here, $A_{\mu}$ and $F_{\mu\nu}$ are obtained from the resultant Hamiltonian (for the definitions, see \cref{appsec: phi A F}).
Vector $\bm{p}$ parametrizes the $3$-sphere in the momentum space. 

We demonstrate the emergence of a symmetry-protected HEP$4$ by analyzing a toy model whose Hamiltonian reads
\begin{align}
\label{eq: toy Hopf EP4}
H &= 
\left(
\begin{array}{cccc}
0 & 1 & 0 & 0  \\
0 & 0 & 1 & 0 \\
0 & 0 & 0 & 1 \\
\frac{\zeta_3}{24^2} & -\frac{\zeta_2}{24} & \frac{\zeta_1}{2} & 0 \\
\end{array}
\right), 
\end{align}
with real functions $\zeta_{1,\ldots,3}(\bm{k})$ parametrized by $\delta$ and $m_0$ [for the explicit form, see \cref{appsec: toy Z EP4}].
As is the case with Hamiltonians in \cref{eq: toy Z2EP3,eq: toy Z2 EP5}, the Hopf topology~\cite{Moore_HopfIns_PRL2008} of $\zeta$'s in the perturbed model [\cref{eq: toy Hopf EP4}] is equivalent to that of the resultant vector.

The vanishing resultant vector 
$\bm{R}\propto(\zeta_1,\zeta_2,\zeta_3 )^{\mathrm{T}}=\bm{0}$ 
specifies where the symmetry-protected HEP4 emerges in the momentum space. 
The explicit form of the resultant [see \cref{appsec: toy Z EP4}] indicates the emergence of a symmetry-protected HEP4 at $\bm{k}\,{=}\,\bm{0}$ for $m_0\,{=}\,3$ [see \cref{fig: HopfEP4_Ek_inv}(a,b)].
Here, we characterize the symmetry-protected HEP4 enclosed by two planes at $k_4\,{=}\,-\pi/2$ and $k_4\,{=}\,\pi/2$.
Numerically evaluating the Hopf invariant [\cref{eq: Hopf inv}], we obtain $\nu_{\mathrm{H}}\,{=}\,0$ [$\nu_{\mathrm{H}}\,{=}\,1$] for $k_4\,{=}\,-\pi/2$ [$k_4\,{=}\,\pi/2$] which is consistent with linking of the inverse maps in the momentum space  [see \cref{fig: HopfEP4_Ek_inv}(c,d)]~\cite{Wilczek_Linking_PRL1983,Kennedy_Hopf_PRB2016}. 
These results indicate that the symmetry-protected HEP4 is characterized by~$\nu_{\mathrm{H}}\,{=}\,1$.

\section{
General characteristics of HEPs
}

\subsection{Multiply-charged aspect of HEPs}
%
Reference~\citenum{Delplace_EP3_PRL2021} has pointed out that the codimension of symmetry-protected EP$4$s is three, whereas the codimension of the HEP$4$ in our model is four [see \cref{fig: HopfEP4_Ek_inv}(a,b)].
This mismatch in codimension implies that the HEP$4$ inflates into a loop of EP$4$s under a generic perturbation of the Hamiltonian matrix [see \cref{fig: HopfEP4_Ek_inv}(e)].
Such a perturbation does not trivialize the Hopf topology but enriches it~\cite{Bzdusek_TSMclass_PRB2017}: the loop of EP$4$s carries both the Hopf invariant on $S^3$ and the resultant winding number on $S^2$.~\footnote{
In contrast to the EP4s with the Hopf topology, a loop of formerly reported EP4s can be contracted and annihilated.
} 
In the same spirit, a similar assignment of multiple topological invariants applies to all HEPs introduced in this work.

\subsection{Additional bands}

So far, we have discussed HEP$n$s in $n$-band models. 
For systems with more than $n$ bands, EP$n$s are still associated with winding of the resultant vector; however, the converse does not hold.
Specifically, as discussed in~\cref{appsec: FakeEP3}, it is possible to find situations where a finite value of the winding number leads to the vanishing resultant vector $\bm{R}\,{=}\,\bm{0}$ without a symmetry-protected~EP3. 

Nevertheless, topological invariants of resultants are applicable when we focus around an HEP$n$ (or an EP$n$).
Specifically, if we know that an HEP$n$ emerges at momenta $\bm{k}\,{=}\,\bm{k}_0$ and energy $E\,{=}\,E_0$, we may apply 
the Taylor expansion to the characteristic polynomial $P(E)=\mathrm{det}[H(\bm{k})-E\1]$, 
which leads to a polynomial $\tilde{P}(\tilde{E})$ with degree $n$
\begin{align}
\label{eq: P tilde}
\tilde{P} (\tilde{E}) &= \sum^n_{j=0} a_j(\bm{k}) \tilde{E}^j,
\end{align}
with $a_j(\bm{k})$ ($j\,{=}\,0,1,\ldots, n$) being complex functions and $\tilde{E}\,{=}\,E-E_0$.
Computing the topological invariant [e.g., \cref{eq: Z2 inv}] of $\tilde{P} (\tilde{E})$ around $\bm{k}\,{=}\,\bm{k}_0$, we can characterize the HEP$n$ for systems with more than $n$ bands. 

\begin{table}[t]
    \centering
    \begin{tabular}{ccccc} 
          \hline\hline
  \multirow{3}{*}{~$c$~}  & ~HEP$3$~                                    & ~\multirow{2}{*}{~SP-HEP$6$~}~                   & ~HEP$4$~                             \\       
                               & ~SP-HEP$4$, SP-HEP$5$~                  &                                                                 & ~SP-HEP$7$~                  \\ \hline 
       ~$4$~                   & ~\tableblue{$\mathbb{Z}$}~                         & ~$0$~                                                           & ~$0$~                                       \\
       ~$5$~                   & ~\tableblue{$\mathbb{Z}_2$}~                       & ~$\mathbb{Z}$~                                                  & ~$0$~                                       \\       
       ~$6$~                   & ~$\mathbb{Z}_2$~                                   & ~$\mathbb{Z}_2$~                                                & ~$\mathbb{Z}$~                              \\ 
       ~$7$~                   & ~$\mathbb{Z}_{12}$~                                & ~$\mathbb{Z}_2$~                                                & ~$\mathbb{Z}_2$~                            \\ 
       ~$8$~                   & ~$\mathbb{Z}_{2}$~                                 & ~$\mathbb{Z}\times \mathbb{Z}_{12}$~                            & ~$\mathbb{Z}_2$~                            \\ 
       ~$9$~                   & ~$\mathbb{Z}_{2}$~                                 & ~$\mathbb{Z}^2_2$~                                              & ~$\mathbb{Z}_{24}$~                         \\ 
       ~$10$~                   & ~$\mathbb{Z}_{3}$~                                 & ~$\mathbb{Z}^2_2$~                                              & ~$\mathbb{Z}_{2}$~                          \\  
       ~$11$~                  & ~$\mathbb{Z}_{15}$~                                & ~$\mathbb{Z}_{24}\times \mathbb{Z}_3$~                          & ~$\mathbb{Z}_{2}$~                          \\  
       ~$12$~                  & ~$\mathbb{Z}_{2}$~                                 & ~$\mathbb{Z}_{15}$~                                             & ~$\mathbb{Z}_{2}$~                          \\  
       ~$13$~                  & ~$\mathbb{Z}^2_{2}$~                               & ~$\mathbb{Z}_2$~                                                & ~$\mathbb{Z}_{30}$~                         \\  
       ~$14$~                  & ~$\mathbb{Z}_{12}\times\mathbb{Z}_{2}$~            & ~$\mathbb{Z}^3_2$~                                              & ~$\mathbb{Z}_{2}$~                          \\ 
       ~$15$~                  & ~$\mathbb{Z}_{84}\times\mathbb{Z}^2_{2}$~          & ~$\mathbb{Z}_{120}\times\mathbb{Z}_{12}\times\mathbb{Z}_{2}$~  & ~$\mathbb{Z}^3_{2}$~                        \\          
       ~$16$~                  & ~$\mathbb{Z}^2_2$~                                 & ~$\mathbb{Z}_{84}\times \mathbb{Z}^5_2$~                               & ~$\mathbb{Z}_{72}\times \mathbb{Z}_{2} $~   \\  \hline\hline
    \end{tabular}
     \caption{
       Hopf exceptional points (HEPs) are classified by higher homotopy groups of spheres.
       The topology explicitly analyzed in this manuscript corresponds to the blue entries. 
       HEP$n$s [symmetry-protected HEP$n$s] of codimension $c\,{=}\,4,5,\ldots$ are classified by $\pi_{c-1}(S^{2n-3})$ [$\pi_{c-1}(S^{n-2})$].
     }
    \label{tab: homotopy table}
\end{table}

\subsection{Higher dimensions}

Higher homotopy groups of spheres indicate abundant topology of HEP$n$s. 
HEP$n$s are captured by the ($2n-2$)-component resultant vector, and thus are classified by $\pi_{c-1}(S^{2n-3})$ with codimension $c \geq 4$. Symmetry-protected HEP$n$s are captured by the ($n-1$)-component resultant vector, and thus are classified by $\pi_{c-1}(S^{n-2})$. 
The classification results for HEP$n$s ($n\,{=}\,3,4$) and symmetry-protected HEP$n$s ($n\,{=}\,4,5,6,7$) are summarized in \cref{tab: homotopy table} (see also  Sec.~4.1 of Ref.~\cite{AHatcher_Textbook2022} and Chapter~XIV of Ref.~\cite{Toda_textbook1962}).
This table predicts the presence of various HEP$n$s following exotic fusion rules. For instance, $\mathbb{Z}_3$ topology of codimension $c\,{=}\,10$ implies an HEP$3$ annihilate with two copies of itself, which is reminiscent of parafermions~\cite{Read_Parafermion_PRB1999,Fendley_Parafermions_JStatMech2012,Mong_PRX2014,Liu_Moire_NatComm2025}. 
In addition, there exist HEP$3$s with $\mathbb{Z}_{12}$ topology of codimension $c\,{=}\,7$ and HEP$4$s with $\mathbb{Z}_{24}$ topology of codimension $c\,{=}\,9$.
Topological invariants for these HEP$n$s remain to be formulated.

\section{
Conclusion and outlook
}
We have discovered novel non-Hermitian topological excitations, dubbed Hopf exceptional points.
Saliently, HEP$3$s and symmetry-protected HEP$5$s exhibit an unusual $\mathbb{Z}_2$ topology, meaning that they act as their own antiparticle. 
This striking feature arises from the homotopy group $\pi_4(S^3)\,{=}\,\mathbb{Z}_2$. 
We have further discovered symmetry-protected HEP4s whose topology is classified by $\pi_3(S^2)\,{=}\,\mathbb{Z}$. 
Leveraging higher homotopy groups of spheres, we elucidate the potential presence of HEP$n$s characterized by abundant finite groups (e.g., $\mathbb{Z}_3$, $\mathbb{Z}_{12}$, and $\mathbb{Z}_{24}$) beyond the classification table of Bernard-LeClair symmetry classes.

Our work on non-Hermitian multiband systems opens up a new direction of topological physics. 
Here, we outline several concrete open questions motivated by our findings.
On one hand, from the implementation perspective, we anticipate that
HEP$n$s can be realized in a variety of experimental settings. Notably, they may be realized in terms of non-unitary photon dynamics where both the eigenvalues and eigenstates have been measured accurately at and around multifold EPs \cite{Wang_EPnPTPhotoSciAdv2023}, hence suggesting a path to a particularly comprehensive simulation of HEP$n$s. Further promising platforms include nitrogen-vacancy spin systems \cite{Wu_EP3openQ_NatNano2024} and coupled micro-resonators \cite{Hodaei_EPnPT_Nature2017} in which multifold EPs have also been realized.
Quite generally, the high controllability of metamaterials allows access to momentum (or parameter) spaces with dimensions larger than three~\cite{Dembowski_EPPhoto_PRL2001,Lee_EPPhoto_PRL2009,Tang_EPn_Science2020,Tang_EP3Mech_NatComm2023,Yoshida_EPn_PRR2024}, thus inviting various realistic experimental verifications of the unusual topology of HEP$n$s.

On the other hand, our work also indicates several concrete theoretical aspects that deserve a deeper mathematical analysis. 
First, \cref{tab: homotopy table} implies the presence of novel HEPs following a unique fusion rule due to their topology. 
Explicit analysis based on topological invariants and concrete models is an interesting issue.
Next, unless an explicit band structure is provided, one-to-one correspondence between EP$n$s and the resultant winding numbers is lost for many-band systems.
This observation calls for a more general topological characterization of Hopf exceptional points.
Finally, it is interesting to consider non-Hermitian topological bands, captured by suitably adapted Hopf invariants~\cite{Kennedy_Hopf_PRB2016}, that arise in models with a hopfion texture~\cite{Rybakov_MagHopfion_APLMat2022} of the resultant vector over the Brillouin zone~torus.
We postpone the analysis of these questions to future studies.

\section*{Acknowledgments} 
T.~Y.~is grateful for the support from the ETH Pauli Center for Theoretical Studies.

\paragraph{Funding information} T.~Y.~is supported by JSPS KAKENHI Grant Nos.~JP21K13850, JP23KK0247, JP25K07152, and JP25H02136 as well as JSPS Bilateral Program No.~JPJSBP120249925 and Yamada Science Foundation.
E.~J.~B.~is supported by the  Wallenberg Scholars program (2023.0256) and the G{\"o}ran Gustafsson Foundation for Research in Natural Sciences and Medicine.
T.~B.~was supported by the Starting Grant No.~211310 by the Swiss National Science Foundation (SNSF).

\begin{appendix}
\numberwithin{equation}{section}

\section{Defining topology from resultants}
\label{appsec: Res-all}

\subsection{Resultant of a pair of polynomials}
\label{appsec: Res}
For given two polynomials 
\begin{align}
 f(x) &= a_nx^n+\ldots+a_1x+a_0, \\
 g(x) &= b_mx^m+\ldots+b_1x+b_0,
\end{align}
with complex coefficients $a$'s and $b$'s, the resultant is defined as 
\begin{align}
 \mathrm{Res}[f(x),g(x)] &= a^m_nb^n_m\prod_{i,j}(\alpha_i-\beta_j).
\end{align}
Here, $\alpha$'s and $\beta$'s are roots of polynomials $f(x)$ and $g(x)$, respectively. 
Symbol $\textstyle{ \prod_{i,j}}$ denotes the product over all pairs of the roots.
The resultant vanishes when the two polynomials have a common root.

The resultant can be computed from the Sylvester matrix 
\begin{align}
\mathrm{Res}[f(x),g(x)] &= \mathrm{det}
    \left[
    \begin{array}{ccccccc}
        a_n & \cdots & a_0    &        &        &  &    \\
            &  a_n   & \cdots & a_0    &        &  &   \\
            &        & \ddots &        & \ddots &  &    \\
            &        &        &    a_n & \cdots & a_0   \\
        b_m & \cdots & b_0    &        &        &       \\
            & b_m    & \cdots & b_0    &        &       \\
            &        & \ddots &        & \ddots &       \\
           &        &        & b_m    & \cdots & b_0   \\
    \end{array}
    \right]. 
\end{align}
Here, the empty elements are zero.
The size of the matrix is $n+m$ and the first $n$ rows are composed of the coefficients $a$'s and the remaining $m$ rows are composed of the coefficients~$b$'s.

\subsection{Resultant winding number}
\label{appsec: ResW}

We next consider an $N$-component resultant vector $\bm{R}(\bm{k})$, such as the one in 
\cref{eq: Rvec Generic 3x3}, defined on an $N$-dimensional momentum space $\mathbb{R}^N$.
When the norm is finite $\lVert\bm{R}\rVert \neq 0$ on an $(N-1)$-sphere $S^{N-1}\subset \mathbb{R}^N$, one can consider the normalized vector $\bm{n}(\bm{k})=\bm{R}(\bm{k})/\lVert\bm{R}(\bm{k})\rVert$ whose topology is classified by $\pi_{N-1}(S^{N-1})=\mathbb{Z}$.

The topology of such a map is characterized by the resultant winding number $W_{N-1}$~\cite{Delplace_EP3_PRL2021,Yoshida_EPn_PRR2024} 
\begin{align}
\label{eq: Res WN-1}
 W_{N-1} &= \frac{\epsilon^{i_1\cdots i_{N} }}{A_{N-1}} \int d^{N-1}\bm{p} f_{i_1\cdots i_N}(\bm{p}), \\
 f_{i_1\cdots i_N}(\bm{p}) &= n_{i_1}\partial_1 n_{i_2}\partial_2 n_{i_3} \cdots \partial_{N-1} n_{i_N},
\end{align}
where the integral is taken over $S^{N-1}$ in the momentum space parameterized by vector $\bm{p}$.
Here, the area of the $(N-1)$-dimensional sphere $A_{N-1}$ is expressed by
\begin{align}
  A_{2l-1} &= \frac{2\pi^l}{(l-1)!},  \\
  A_{2l-2} &= \frac{ 2^{2l-1} \pi^{l-1} (l-1)!   }{ (2l-2)! },
\end{align}
with a positive integer $l$.
We note that the resultant topology protecting the EP$n$ is neither point-gap nor line-gap topology. 
In general, the EP$n$ is accompanied by manifolds of EP$m$s with $2 \leq m<n$ (e.g., a symmetry-protected EP3 emerges on lines of symmetry-protected EP2s). As EP$m$s require the point-gap closing on the sphere enclosing the EP$n$, the point-gap topology~\cite{Gong_class_PRX18,Kawabata_gapped_PRX19,Zhou_gapped_class_PRB19,Kawabata_EPClass_PRL19}
cannot protect the EP$n$. In addition, the line-gap topology does not protect band touchings for both real- and imaginary parts~\cite{Kawabata_gapped_PRX19}, simultaneously.

The invariants in \cref{eq: Res WN-1} are rewritten as Chern numbers or winding numbers of the resultant Hamiltonian~\cite{Delplace_EP3_PRL2021}. 
Specifically, for $N=3$, \cref{eq: Res WN-1} is rewritten as the first Chern number of the resultant Hamiltonian $H_{\mathrm{R}}(\bm{k})=\bm{R}\cdot\bm{\sigma}$.
For $N=4$, \cref{eq: Res WN-1} is rewritten as the winding number of three-dimensional chiral symmetric Hamiltonian $H_{\mathrm{R}}=\sum_{i=1,\ldots,4} R_i\gamma_i$ satisfying $\gamma_5 H_{\mathrm{R}} \gamma_5 =-H_{\mathrm{R}}$  with $\bm{\gamma}=(\sigma_1\tau_0,\sigma_2\tau_0,\sigma_3\tau_1,\sigma_3\tau_2,\sigma_3\tau_3)^{\mathrm{T}}$. Here, $\tau_{1,\ldots,3}$ are Pauli matrices, and $\tau_0$ is the $2\times 2$ identity matrix.
We note that the presence of the gap of $H_{\mathrm{R}}$ on the sphere is reduced to $\lVert\bm{R}\rVert \neq 0$. This is because the eigenvalues are given by $E_{\mathrm{R}}=\pm \lVert\bm{R}\rVert$ which arises from the anti-commutation relation $\{ \gamma_i, \gamma_j\}=2\delta_{ij}$. 

\subsection{Berry connection and curvature of resultant Hamiltonians for Hopf topology}
\label{appsec: phi A F}

The $\mathbb{Z}_2$ invariant in \cref{eq: Z2 inv} is obtained from the resultant vector $\bm{R}\,{=}\,(R_1,R_2,R_3,R_4)^{\mathrm{T}}$~($R_{1,\ldots,4}\in \mathbb{R}$).
Specifically, the Berry connection $A_{\mu}$ and the Berry curvature $F_{\mu\nu}$ are defined as 
\begin{align}
\label{eq: A}
A_{\mu} &=\frac{1}{2\pi \ii} \langle z  | \partial_{\mu} z \rangle, \\
\label{eq: F}
F_{\mu\nu}&=\frac{1}{2\pi\ii}\Big(\langle \partial_\mu z |\partial_\nu z \rangle - \langle \partial_\nu z |\partial_\mu z \rangle\Big),
\end{align}
with the negative eigenstate $|z \rangle$ of the resultant Hamiltonian $\bm{\tilde{n}}\cdot \bm{\sigma}$ with $\bm{\tilde{n}}\,{=}\,\bm{\tilde{R}}/\lVert\bm{\tilde{R}}\rVert$ and $\bm{\tilde{R}}\,{=}\,(R_1,R_2,R_3)$.
Here, $\sigma_{1,\ldots,3}$ denote Pauli matrices. 
The ratio $\lVert\bm{\tilde{R}}\rVert/R_4$ defines the phase $\Delta\varphi(\bm{p})\,{=}\,2\varphi(\bm{p})$ with $\varphi\,{=}\,\arctan \lVert\bm{\tilde{R}}\rVert/R_4$ ($0 \leq \varphi \leq \pi$). 
Substituting $\Delta\varphi$, $A_{\mu}$, and $F_{\mu\nu}$ into \cref{eq: Z2 inv} yields the $\mathbb{Z}_2$ invariant $\nu_{\mathrm{F}}$.

The $\mathbb{Z}$ invariant in \cref{eq: Hopf inv} is also obtained from the resultant vector $\bm{R}\,{=}\,(R_1,R_2,R_3)^{\mathrm{T}}$ ($R_{1,\ldots,3}\in \mathbb{R}$).
Specifically, the Berry connection $A_{\mu}$ and the Berry curvature $F_{\mu\nu}$ are defined in the same way as \cref{eq: A,eq: F} except for the definition of $|z \rangle$. 
In this case, $|z \rangle$ is defined as the negative eigenstate of the resultant Hamiltonian $\bm{n}\cdot \bm{\sigma}$ with $\bm{n}\,{=}\,\bm{R}/\lVert\bm{R}\rVert$.
Substituting the specific form of the Berry curvature $A_{\mu}$ and the Berry connection $F_{\mu\nu}$ into \cref{eq: Hopf inv} yields $\mathbb{Z}$ invariant $\nu_{\mathrm{H}}$.

\section{
Details of the presented Hamiltonians
}

\subsection{Model in \texorpdfstring{\cref{eq: toy Z2EP3}}{(4)}
and computation of the \texorpdfstring{$\mathbb{Z}_2$}{Z2} invariant
}
\label{appsec: toy Z2EP3}
The explicit form of $\zeta$'s is 
\begin{subequations}
\label{eq: zeta Z2EP3}
\begin{align}
 \zeta_1 &= -2 \sin\phi \Big( \eta^*_\uparrow\eta_\downarrow \Big), \\
 \zeta_2 &=-
 \sin\phi \Big(|\eta_\uparrow|^2-|\eta_\downarrow|^2\Big)+\ii \cos \phi.
\end{align}
\end{subequations}
Here $\eta_\uparrow$,  $\eta_\downarrow$ and $\phi$ are defined as
\begin{subequations}
\label{eq: eta and phi Z2EP3}
\begin{align}
\eta_\uparrow &= \sin k_1+\ii f(k_5)\sin k_2,\\
\eta_\downarrow &= \sin k_3+\ii \Big[\xi(\bm{k})+\frac{3}{2}\sin k_4 -3(\cos k_5+\delta ) \Big], \\
\phi& =\frac{\pi}{2} (1-\cos k_4), 
\end{align}
\end{subequations}
with $\xi(\bm{k})\,{=}\,\sum_{j=1,2,3}\cos k_j-m_0$ and $f(k_5)$ being $f(k_5)\,{=}\,1$ or $f(k_5)\,{=}\,2\sin (k_5/2)$.

The resultant vector $\bm{R}$ is obtained as 
\begin{align}
\label{eq: toy Rvec Z2EP3}
\bm{R} 
       &= (3!)^2 \Big(-2\sin\phi \mathrm{Re}[\eta_\uparrow\eta_\downarrow],
       -2\sin\phi \mathrm{Im}[\eta_\uparrow\eta_\downarrow], -\sin\phi (|\eta_\uparrow|^2-|\eta_\downarrow|^2),
       \cos\phi 
\Big)^\mathrm{T}.
\end{align}
The above equation indicates that the resultant vector vanishes when $\eta_{\uparrow}\,{=}\,\eta_{\downarrow}\,{=}\,0$ and $\phi\,{=}\,\pi/2$ both hold.

From the given resultant vector [see \cref{eq: toy Rvec Z2EP3}], the $\mathbb{Z}_2$ invariant $\nu_\mathrm{F}$ is obtained as follows.
With $\bm{\tilde{n}}\,{=}\,\bm{\tilde{R}}/\lVert\bm{\tilde{R}}\rVert$ and \cref{eq: toy Rvec Z2EP3}, the resultant Hamiltonian is obtained as 
\begin{align}
& \tilde{\bm{n}}\cdot\bm{\sigma}= -\frac{1}{\sqrt{|\eta_\uparrow|^2+|\eta_\downarrow|^2}}
\!\left(
\begin{array}{cc}
\!\! |\eta_\uparrow|^2-|\eta_\downarrow|^2 & 2\eta_\uparrow \eta^*_\downarrow \\
2\eta^*_\uparrow \eta_\downarrow & -|\eta_\uparrow|^2+|\eta_\downarrow|^2 \!\!
\end{array}
\right)\!.
\end{align}
Thus, the negative eigenstate of $\tilde{\bm{n}}\cdot\bm{\sigma}$ is obtained as
\begin{align}
|z(\bm{k})\rangle \,{=}\, \frac{1}{\sqrt{|\eta_\uparrow|^2+|\eta_\downarrow|^2}}
\left(
\begin{array}{c}
\eta_\uparrow  \\
\eta_\downarrow 
\end{array}
\right).
\end{align}
%
In addition, from the resultant vector in \cref{eq: toy Rvec Z2EP3}, we obtain 
\begin{align}
\Delta\varphi=2\arctan \lVert \tilde{\bm{R}} \rVert / R_4=2\phi.
\end{align}
Substituting the obtained $|z\rangle$ and $\Delta \varphi$ into \cref{eq: Z2 inv}, we can numerically compute the $\mathbb{Z}_2$ invariant.

\subsection{Model in \texorpdfstring{\cref{eq: toy Z2 EP5}}{Eq. (8)}}
\label{appsec: toy Z2EP5}
%
The explicit form of $\zeta$'s are 
\begin{subequations}
\label{eq: zeta Z2 PTEP5}
\begin{align}
& \zeta_1+\ii \zeta_2 = -2 \sin\phi(\eta_\uparrow\eta_\downarrow), \\
& \zeta_3 = -
 \sin\phi (|\eta_\uparrow|^2-|\eta_\downarrow|^2), \\
& \zeta_4 = \cos \phi,
\end{align}
\end{subequations}
with $\eta$'s and $\phi$ defined in Eq.~\eqref{eq: eta and phi Z2EP3}.

The resultant vector of this Hamiltonian is obtained as
\begin{align}
\label{eq: toy Rvec Z2EP5}
\bm{R} &=  (5!)^2 \Big(-2\sin\phi \mathrm{Re}[\eta_\uparrow\eta_\downarrow],
       -2\sin\phi \mathrm{Im}[\eta_\uparrow\eta_\downarrow], -\sin\phi (|\eta_\uparrow|^2-|\eta_\downarrow|^2),
       \cos\phi 
\Big)^\mathrm{T},
\end{align}
which is proportional to the resultant vector in Eq.~\eqref{eq: toy Rvec Z2EP3}.
The above equation indicates that the resultant vector vanishes when $\eta_{\uparrow}\,{=}\,\eta_{\downarrow}\,{=}\,0$ and $\phi\,{=}\,\pi/2$ both hold.
%

\subsection{Model in \texorpdfstring{\cref{eq: toy Hopf EP4}}{Eq. (13)}}
\label{appsec: toy Z EP4}
The explicit form of $\zeta$'s are 
\begin{subequations}
\label{eq: zeta ZEP4}
\begin{align}
& \zeta_1+\ii \zeta_2 = 2(\eta^*_\uparrow\eta_\downarrow), \\
&\zeta_3
=
|\eta_\uparrow|^2-|\eta_\downarrow|^2+\delta.
\end{align}
\end{subequations}
Here $\eta_\uparrow$ and $\eta_\downarrow$ are 
\begin{subequations}
\begin{align}
\label{eq: eta up 3D}
\eta_\uparrow &= \sin k_1+\ii \sin k_2, \\
\label{eq: eta down 3D}
\eta_\downarrow &= \sin k_3+\ii \Big[\xi(\bm{k})+\sin k_4 \Big],
\end{align}
\end{subequations}
with $\xi(\bm{k})\,{=}\,\sum_{j=1,2,3}\cos k_j-m_0$.

The resultant vector $\bm{R}$ is obtained as 
\begin{align}
\label{eq: toy Rvec ZEP4}
 \bm{R}&=
-(24)^2\Big(
2\mathrm{Re}[\eta^*_\uparrow\eta_\downarrow],
2\mathrm{Im}[\eta^*_\uparrow\eta_\downarrow],
|\eta_\uparrow|^2-|\eta_\downarrow|^2+\delta \Big)^{\mathrm{T}}.
\end{align}
The above equation indicates that the resultant vector vanishes when $\eta_\uparrow\,{=}\,0$ and $|\eta_{\downarrow}|\,{=}\,\sqrt{\delta}$ both hold for $\delta>0$ or when $\eta_\downarrow\,{=}\,0$ and $|\eta_{\uparrow}|\,{=}\,\sqrt{|\delta|}$ both hold for $\delta<0$.

\section{
Fake EP3s
}
\label{appsec: FakeEP3}

For three-band systems, EP3s with $PT$-symmetry are characterized by the resultant winding number~\cite{Delplace_EP3_PRL2021,Yoshida_EPn_PRR2024} of the resultant Hamiltonian $R_1\sigma_1+R_2\sigma_2$ with  $\bm{R}=(r_1,r_2)$ which satisfies chiral symmetry. However, for systems with four or more bands, the resultant winding number may additionally capture ``fake EP3s" unless Taylor expansion is applied.
This is because the resultant vector may vanish without a triple root in the characteristic polynomial $P(E)$ whose degree is four or higher.

\begin{figure}[!t]
\begin{center}
\includegraphics[width=0.75\hsize,clip]{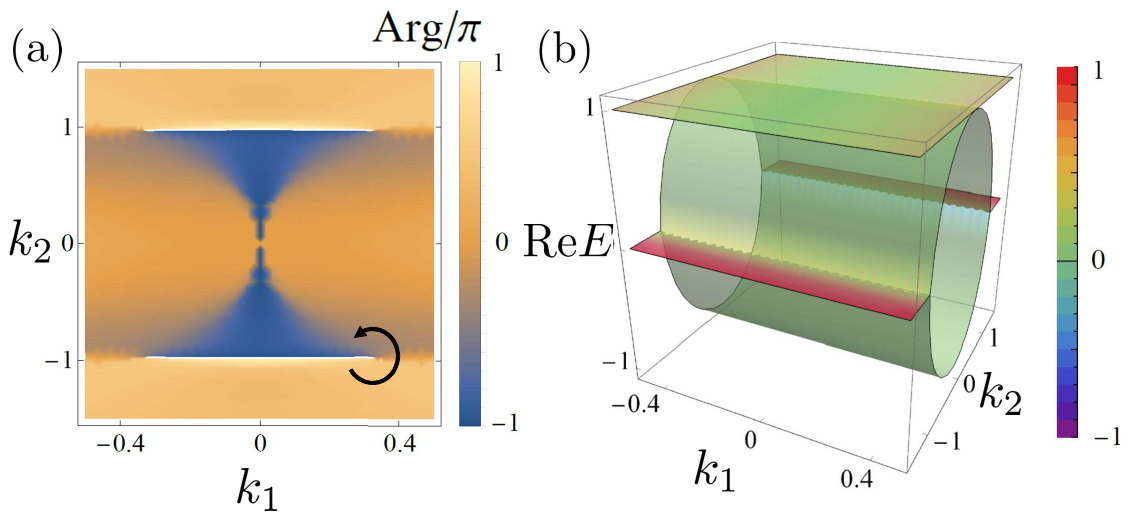}
\end{center}
\caption{
(a) [(b)]:~ Argument of $R_1+\ii R_2$ [band structure] of Hamiltonian~\eqref{eq: toy H fake EP3}. 
In panel (b), the complex conjugate of the top band is omitted.
}
\label{fig: FakeEP3}
\end{figure}

As an example, we consider a non-Hermitian Hamiltonian 
\begin{align}
\label{eq: toy H fake EP3}
H &= 
\left(
\begin{array}{cccc}
1+\ii k_1 & 0 & 0 & 0  \\
0 & 1-\ii k_1 & 0 & 0 \\
0 & 0 & \ii k_2 & 1 \\
0 & 0 & 1 & -\ii k_2 \\
\end{array}
\right),
\end{align}
satisfying $PT$-symmetry [\cref{eq: pT symm}] with 
\begin{align}
U_{\mathrm{PT}} &= 
\left(
\begin{array}{cccc}
0 & 1 & 0 & 0  \\
1 & 0 & 0 & 0 \\
0 & 0 & 0 & 1 \\
0 & 0 & 1 & 0 \\
\end{array}
\right).
\end{align}
%
The resultant winding number is defined as
\begin{align}
W_1 &= \oint \frac{dp}{2\pi \ii} \partial_p \log [R_1(p)+\ii R_2(p)],  
\end{align}
where $p$ parametrizes the circle in the momentum space. 
The winding number $W_1$ (see \cref{appsec: ResW}) is finite for the path illustrated by the black arrow in \cref{fig: FakeEP3}(a). However, the system does not host EP3s [see \cref{fig: FakeEP3}(b)]. 

\end{appendix}




\end{document}